\documentclass{nseJournal}

\begin{document}

\title{On the use of graph theory to interpret the output results from a Monte-Carlo depletion code}

\addAuthor{\correspondingAuthor{Benjamin Dechenaux}}{a}
\correspondingEmail{benjamin.dechenaux@irsn.fr}

\addAffiliation{a}{Institut de Radioprotection et de Sûreté Nucléaire\\ 31 avenue de la division Leclerc, 92260 Fontenay-aux-Roses, France}

\addKeyword{Graph theory}
\addKeyword{Monte Carlo}
\addKeyword{Depletion}
\addKeyword{VESTA}

\titlePage

\begin{abstract}
The analysis of the results of a depletion code is often considered a tedious and delicate task for it requires both the processing of large volume of information (the time dependent composition of up to thousands isomeric states) and an extensive experience of nuclear reactions and associated nuclear data. 
From these observations, dedicated developments have been integrated to the upcoming version of the Monte Carlo depletion code \textsc{VESTA}~2.2, in order to implement an innovative representation of depletion problems. The aim is to provide user with an adapted and efficient framework to ease the analysis of the results of the code and facilitate their interpretation. This effort ultimately culminated in the development of the representation of the isotope evolution of a given system as a directed graph.  

In this paper, it is shown that the Bateman equation encoded in the \textsc{VESTA} code indeed possesses a natural interpretation in terms of directed cyclic graph and it is proposed to explore some of the insight one can gain from the graph representation of a depletion problem.
Starting from the new capabilities of the code, it is shown how one can build on the wealth of existing methods of graph theory in order to gain useful information about the nuclear reactions taking place in a material under irradiation. 
The graph representation of a depletion problem being especially simple in activation problems, for then only a limited number of nuclides and reactions are involved, the graph representation and its associated tools will be used to study the evolution of the structure materials of a simplified model of the ITER fusion reactor. 
\end{abstract}

\section{Introduction}

\begin{figure}[!ht]
\centering
\includegraphics[scale=0.7]{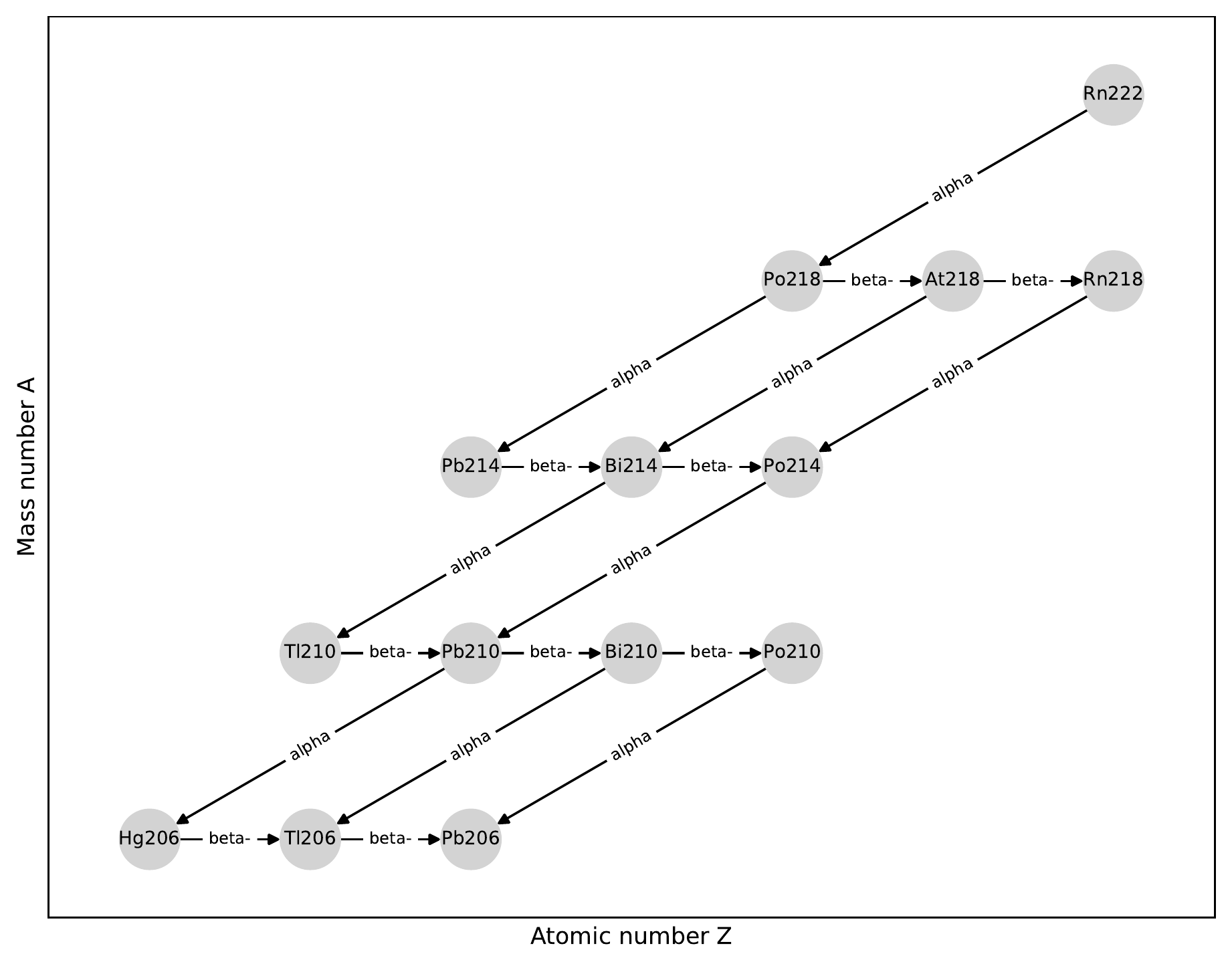}
\caption{The decay chain of $^{222}$Rn. \label{GraphDecayExample}}
\end{figure}

A graph is undoubtedly the most natural and convenient way to pictorially represent the interaction existing between different objects. As such, it has long been in use in a wide variety of domains, with application to the modelling of social interactions, road network studies, internet mapping and many more \cite{NewmannNetworks}. 
In the field of nuclear science, a graph can be used to visually represent the web of nuclear reactions occurring between nuclides. It is indeed the natural way nuclear scientists picture nuclear reaction chains, by imagining a set of arrows pointing from one isotope to another, to form an entangled web of nuclear reactions. An example of such system is displayed on Figure \ref{GraphDecayExample} and represent the graph associated with the decay chain of $^{222}$Rn, as extracted from the JEFF-3.3 nuclear data library \cite{jeff33}. The graph is built using the NetworkX \cite{networkx} Python library.

Although useful, this type of representation seems to reach its limits when the number of nuclides and reactions involved increases, since the intricacy and number of nuclear reactions and nuclides to keep track of makes any visual representation too complicated. 
This situation is often encountered by users of fuel depletion codes, such as VESTA \cite{VESTA}, where the numbers of nuclear reactions and  isotopes involved make the processing of the results of the code a complex and delicate matter.

This complexity might certainly be the main reason why graph theory never totally made its way through the domain of nuclear science (the only use of graph theory in the context of fuel depletion problem that the author is aware of can be found in \cite{fispactConf}). But there is much more to graph theory than a mere visual representation problem. Graph theory effectively come equipped with powerful and systematic tools to process and analyse large amount of data. The developments presented in this paper proposes to explore some of those tools and hope to demonstrate their usefulness in the context of realistic depletion problems.

Another reason for the limited use of graph theory in nuclear science might be the availability of efficient and user friendly tools to manipulate graph objects. The analysis proposed in this paper made extensive use of the NetworkX \cite{networkx} Python package and the Gephi \cite{gephi} graph visualisation and analysis software. 

\section{Graph representation of a depletion problem}

\subsection{Preliminary definitions}
In mathematical terms, a graph $G$ is whatever object which is given by the specification of two particular sets :
\begin{itemize}
    \item a set of \textit{vertices} $V$;
    \item a set of \textit{edges} $E$, being a subset of $V^2$ : $E \subseteq \{ (x,y) | x,y \in V^2\}$.
\end{itemize}

It is clear from Figure \ref{GraphDecayExample} that edges in the graph will be used to encode for nuclear reactions occurring between nuclides, themselves represented as vertices. But, because nuclear reactions are always occurring in an unique direction (i.e. from a source to a target nuclide), one needs to specify that each pair $(x,y)$ of the edge set is to be understood as an \textit{ordered} pair, that is to say that the pairs $(x,y)$ and $(y,x)$ must be viewed as two distinct elements. When this additional structure is given, the underlying graph is said to be \textit{directed}.

Note also that in the definition of the set of edges $E$, nothing prevent the appearance of edges of the form $(x,x)$, i.e. "loops" starting and ending at the same vertex. If, on the model of figure \ref{GraphDecayExample}, each vertex is to represent a specific isotope of the system, then loops might be used to represent isomeric transitions occurring when an isotope is produced in an excited state. In depletion problems, because isomeric states can have a non-negligible impact on the behaviour of the system, they must however be treated as being completely separate entity than their associated ground states. For this reason, a different vertex must be associated in the graph with each isomeric state of a nuclide. Working at this isomeric state level, there is no notion of vertex possessing self loops. In mathematical jargon, graph that do not present self loops are called \textit{simple} graphs.

Finally, numerical values can be assigned to vertices or edges. We then say that the graph is \textit{weighted}. In the practical application of graph theory to depletion problems, it appears natural to assign to each vertex the composition of the isomeric state to which it is associated and to each edge the reaction rate of the associated nuclear reaction. With this additional structure, two different visions of the problem can be envisaged. The first is to consider the problem as flows being exchanged between objects on the fixed, pre-existing grid of all nuclides and possible nuclear reactions. For most purposes, and for visualisation and analysis purposes in particular, a second point of view must be favored and which is to envisage the graph itself as a dynamical entity. As the material composition evolve in time, new vertices and edges will appear or disappear in the graph. 
\newline

The main idea underlying this paper is to try to extract as much information as possible about the behaviour of a material under irradiation by conscientiously analysing snapshots of the simple directed graph representation of the system at different moment in time. 

\subsection{Monte-Carlo fuel depletion simulation with VESTA 2.2}

VESTA \cite{VESTA} is a code whose object is to provide an efficient, accurate and user friendly interface between Monte-Carlo codes used for neutron transport in matter, such as MCNP 6 \cite{MCNP} or MORET 5 \cite{MORET}, and fuel depletion modules such as ORIGEN \cite{ORIGEN}, FISPACT-II \cite{FISPACT} or PHOENIX (VESTA built-in depletion module). This structure makes VESTA a robust and efficient tool to model the evolution of materials in complex geometries and make the code applicable in a wide variety of contexts, from the modeling of commercial BWR and PWR fission reactors to research facilities, fusion devices and many others. 

A typical VESTA simulation consists of an iterative process, where first a simulation of the transport of neutrons is performed. This Monte-Carlo simulation is used to determine the reaction rates of the neutron induced reactions that occur in materials being simulated. This information is then used in a second step that solves the Bateman equations encoding the evolution of the materials' isotopic inventories.    

The results of a typical simulation are the material compositions after each depletion step. These concretely come in the form of lists of concentrations, representing a vast amount of information that, as stated in introduction, is generally difficult to process. The main reason being that when analysing such data, one is left unequipped to disentangle the intricacy of the nuclear reactions occurring in between nuclides and so making sense of where do nuclides come from or disappear is a complicated matter that generally requires knowledge of nuclear reactions and a solid experience in depletion problems. Note that even if it is possible in codes such as VESTA to ask for information about any specific nuclear reaction occurring in the system, an individual treatment has to be performed on each of the thousands of nuclear reaction to fully embrace the complexity of the problem. 
From these observations, some efforts were put and are still ongoing in the development of the latest version 2.2 of the VESTA code and its associated AURORA 1.4 \cite{AURORA} analysis software, in order to provide the user with powerful and systematic ways to ease the processing of depletion output results. 

First, new output files are now provided by VESTA. They allow the user to obtain, at the beginning of each time-step of the depletion process, the reaction rates of each nuclear reaction occurring in the system. The new output, used in conjunction with composition data, is designed to ease the building of the directed graph of the system. Output files can directly be imported in the AURORA analysis software and be used either to export data in a format compatible with most of graph theory tools or directly to analyse graphs within the interface, via a coupling with the Gephi graph visualisation software. 

As already stated, being able to visualise data in the form of a graph is not in itself a sufficient condition to gain insight about a simulation result. But now all of the machinery behind graph theory can be deployed to our advantage. In the following, it is proposed to explore some of the tools that can be employed to analyse results from a typical depletion application example. 

\section{An application example in a fusion activation problem}

To illustrate the usefulness of graph theory as a tool to study output results from a depletion calculation, it is proposed to apply this technique in a concrete application example. To this aim, a very simplified toy model of the ITER fusion device is presented in this section. This model we will be used to simulate the activation of materials that compose a section of the torus of the ITER reactor. 

\begin{figure}[!ht]
\centering
\includegraphics[scale=0.6]{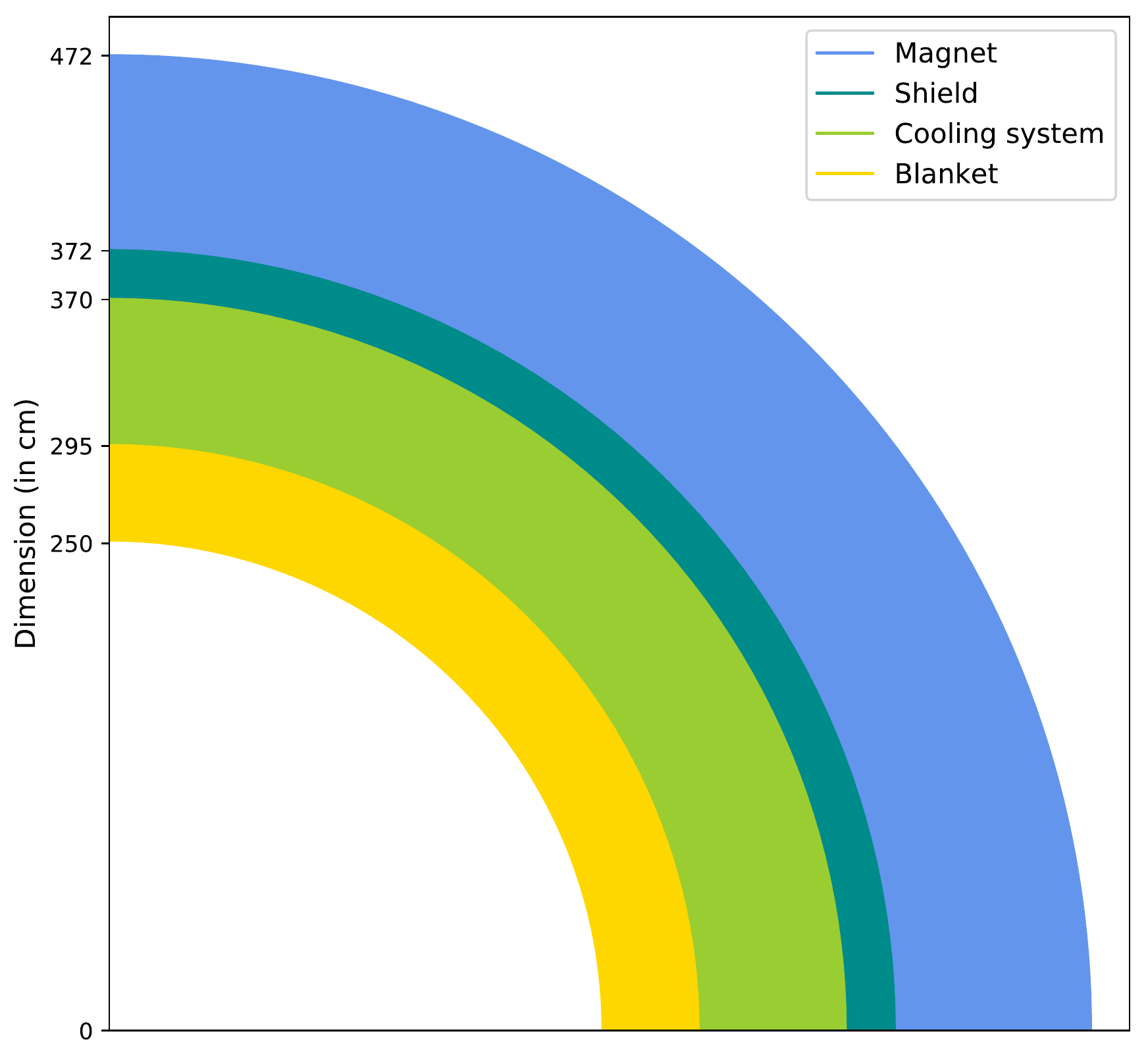}
\caption{One fourth view of the geometric model (dimensions not to scale).\label{ModelGeometry}}
\end{figure}

The model comprises concentric, cylindrical and homogeneous layers of materials. The cylinders are one-meter long and reflection conditions are applied to each of their ends. The model is therefore made, for neutron transport purposes, into an infinite cylinder. A picture of one fourth of the model is shown on Figure \ref{ModelGeometry}. All of the dimensions and compositions in the model are extracted and adapted from \cite{ITERModel}. 

It is assumed in the following that a fusion plasma is continuously sustained for a whole year in the vacuum chamber of the torus, which corresponds to the innermost part of the system. For irradiation purposes, it is supposed that the plasma can be viewed as a constant source of 14 MeV neutrons, emitted along the cylindrical symmetry axis of the system with homogeneous angular dependence. Considering a nominal plasma power of 500 MW carried away exclusively by 14 MeV neutrons, an estimate of $2.10^{20}$ neutrons emitted by the plasma each second is obtained.  

In the following, we will focus our attention on the first, innermost material layer of the model (i.e. the blanket layer shown in yellow on figure \ref{ModelGeometry}). The precise isotopic inventory of the blanket before irradiation is given on Table \ref{tab:params}.

\begin{table}[!h]
\centering
\caption{Initial composition of the blanket (natural abundances).}
\label{tab:params}
\begin{tabular}{|c|c|}
\hline
Isotope & composition  \\
        &(atom/barn/cm) \\
\hline
H & $1.34 \, 10^{-2}$ \\
B & $4.18 \, 10^{ -3}$ \\
N & $3.03 \, 10^{-5}$ \\
O & $6.71 \, 10^{-3}$\\
Si & $2.11\, 10^{-4}$\\
P  & $8.97\, 10^{-6}$\\
Si & $6.35\, 10^{-6}$\\
Cr & $9.33\, 10^{-3}$\\
Mn & $8.90\, 10^{-4}$\\
Fe & $2.86\, 10^{-2}$\\
Co & $2.56\, 10^{-5}$\\
Ni & $7.71\, 10^{-3}$\\
Cu & $3.49\, 10^{-3}$\\
Nb & $3.13\, 10^{-6}$\\
\hline
\end{tabular}
\end{table}

\begin{figure}[!ht]
\centering
\includegraphics[scale=0.6]{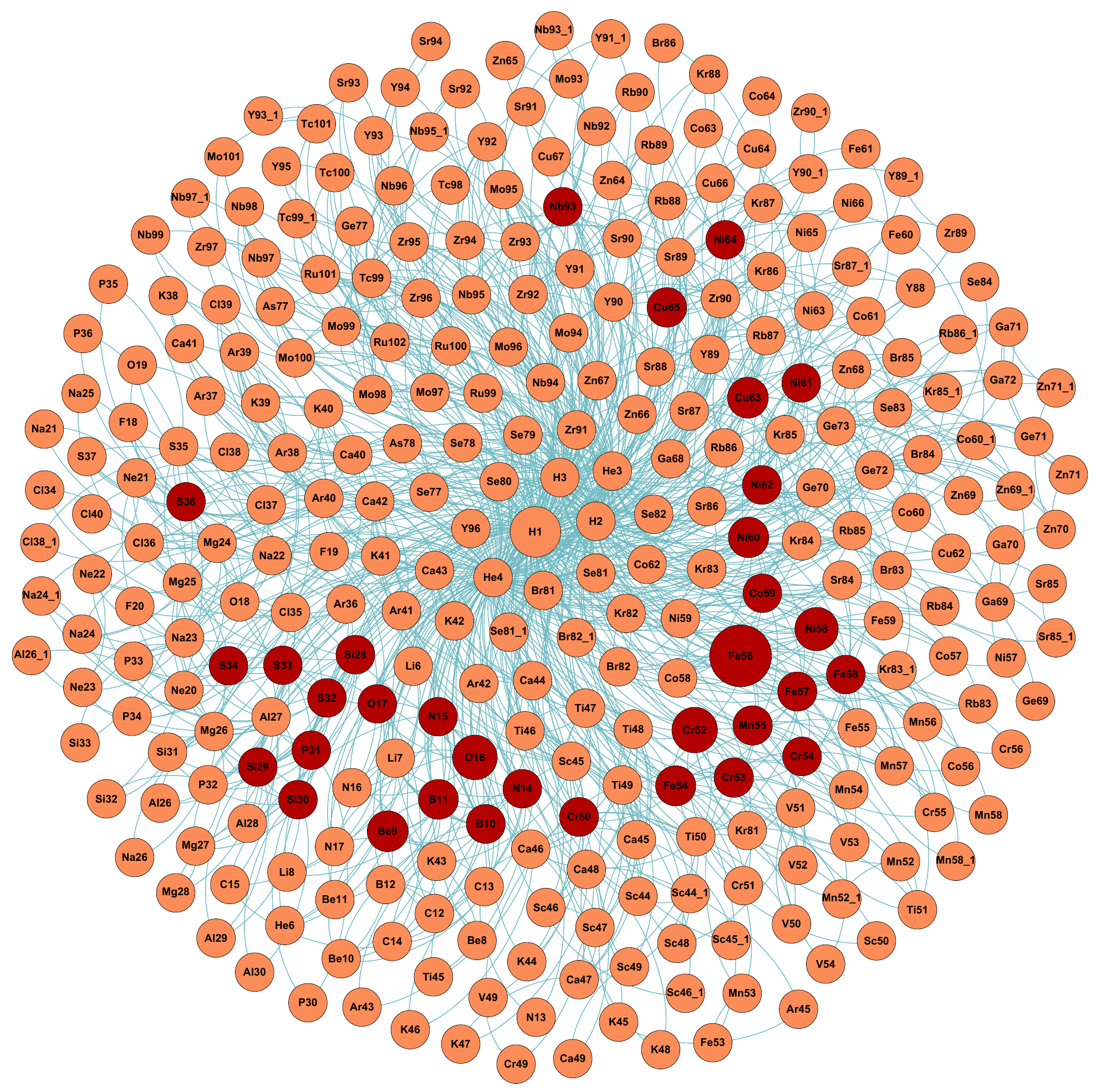}
\caption{Composition of the blanket after 5 days of irradiation.  \label{GraphFusion_initial}}
\end{figure}

A VESTA simulation of the system is performed using the nuclear data library JEFF-3.1 \cite{jeff31}. The simulation is based on MCNP 6.1 for the fixed source neutron transport simulation and the PHOENIX depletion module for simulating the evolution of the composition of the material with time. The composition and (instantaneous) nuclear reaction rates in the material are extracted at different time intervals and are analysed using the graph representation described in introduction. 
Figure \ref{GraphFusion_initial} shows the graph representing the evolved composition of the blanket after five days of irradiation. The graph is made of 309 vertices, representing each isotope and theirs associated isomeric states in the system and the 1531 nuclear reactions (directed edges) occurring in between them. 

\section{Properties extracted from the topological structure of the graph}  

In this section, a focus on the graph of Figure \ref{GraphFusion_initial} is proposed. The graph represent the state of the system after five days of irradiation. Vertices in red represent isotopes that were initially present in the system while vertices colored in orange represent isotopes that were created by the neutron irradiation. Edges in blue represent the instantaneous web of nuclear reactions. Vertices are placed on the graph using the Fruchterman-Reingold force directed algorithm. Finally, here and thorough the rest of this paper, quantities suffixed by an underscore (e.g. "Co60\_1") are to be understood as isomeric states of an original nuclide.

To analyse the graph of the system, it is proposed to use first the properties that are associated with the global "topological" structure of the graph. The goal pursued in this section is to exploit the basic topological structure of the graph to our advantage, reducing when possible the dimensionality of the problem.  

\subsection{Strongly and weakly connected components}

One of the first basic question one could ask when dealing with a graph of any sort as to do with the number of separate pieces it contains. A graph can indeed be constituted by different pieces or \textit{connected components} that does not interact with one another. For undirected graphs (i.e. no notion of direction is defined on the edges), this has a straightforward meaning and the number of components simply reflect the fact that the object is made up of separate independent pieces that can be analysed separately.
The question is more subtle when working with directed graphs, for which connectedness can bear two separate meanings, embodied in the notions of \textit{strong} and \textit{weak} components of a graph.
Weakly connected components of a directed graph correspond to the intuitive notion of separate pieces that the graph is made of. But when the graph is directed, a situation might arise where regions (i.e. subsets of vertices) can be accessed by following a path along the edges of the graph but cannot be left because there is no directed way out of the subset of vertices. In this latter situation, the subset of vertices is said to form a \textit{strongly connected component} of the graph. 

\begin{figure}[!ht]
\centering
\includegraphics[scale=0.6]{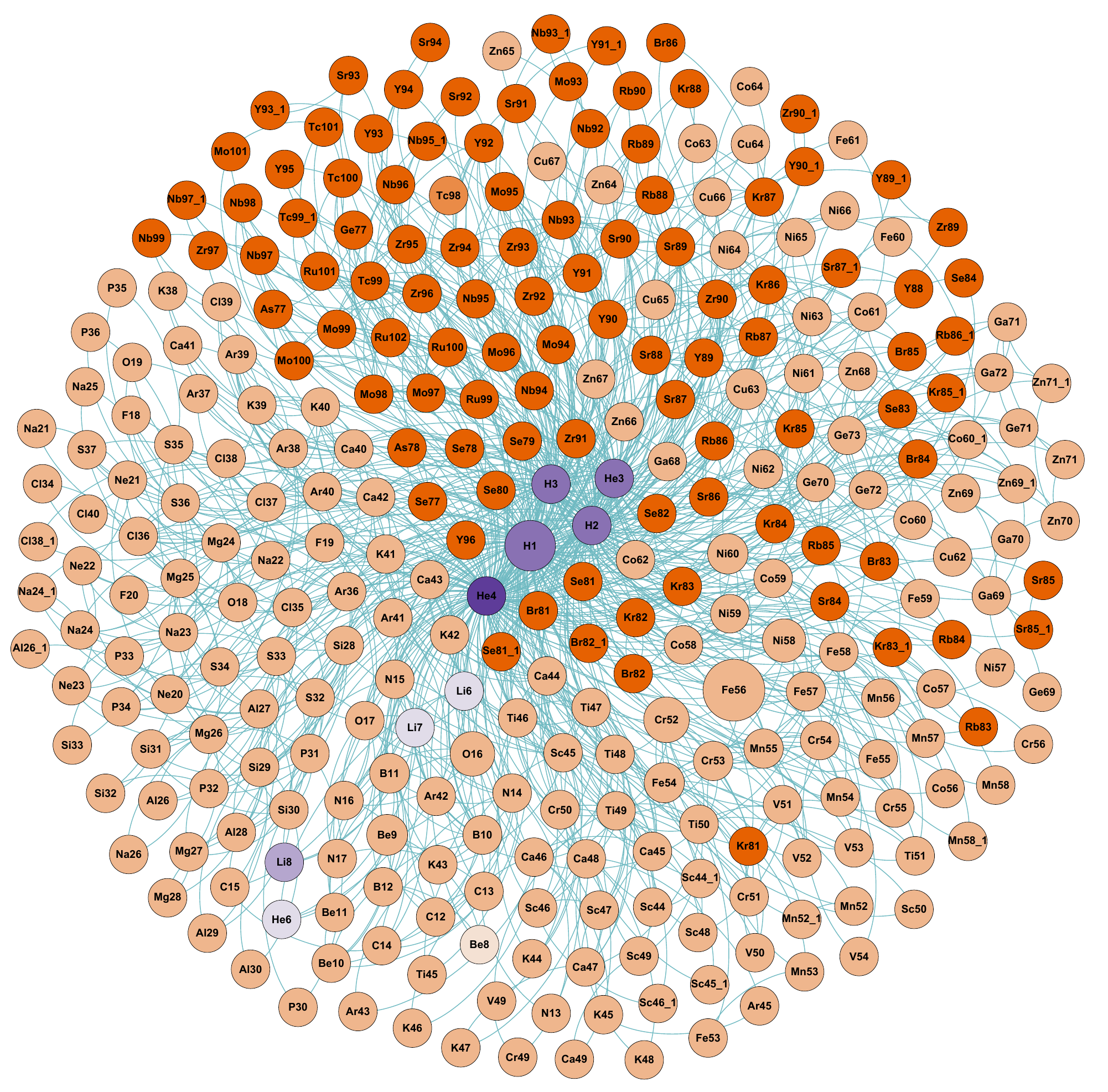}
\caption{Each nuclide on this figure is colored based on the strongly connected component to which it belongs. \label{strongComp}}
\end{figure}

The graph of Figure \ref{GraphFusion_initial} only has one weakly connected component, i.e. the graph of the system comes in one piece. But it has 8 strongly connected components, which are highlighted on Figure \ref{strongComp}. In this figure, it is observed that the graph appears to be built mainly around three of the strongly connected components. All of the other strongly connected components are composed of "dead ends", i.e. stable nuclides. Stable nuclides indeed satisfy the property that they can be reached along a path but can never be left. They therefore form strongly connected components of "unit size". 

\begin{figure}[!ht]
\centering
\includegraphics[scale=0.3]{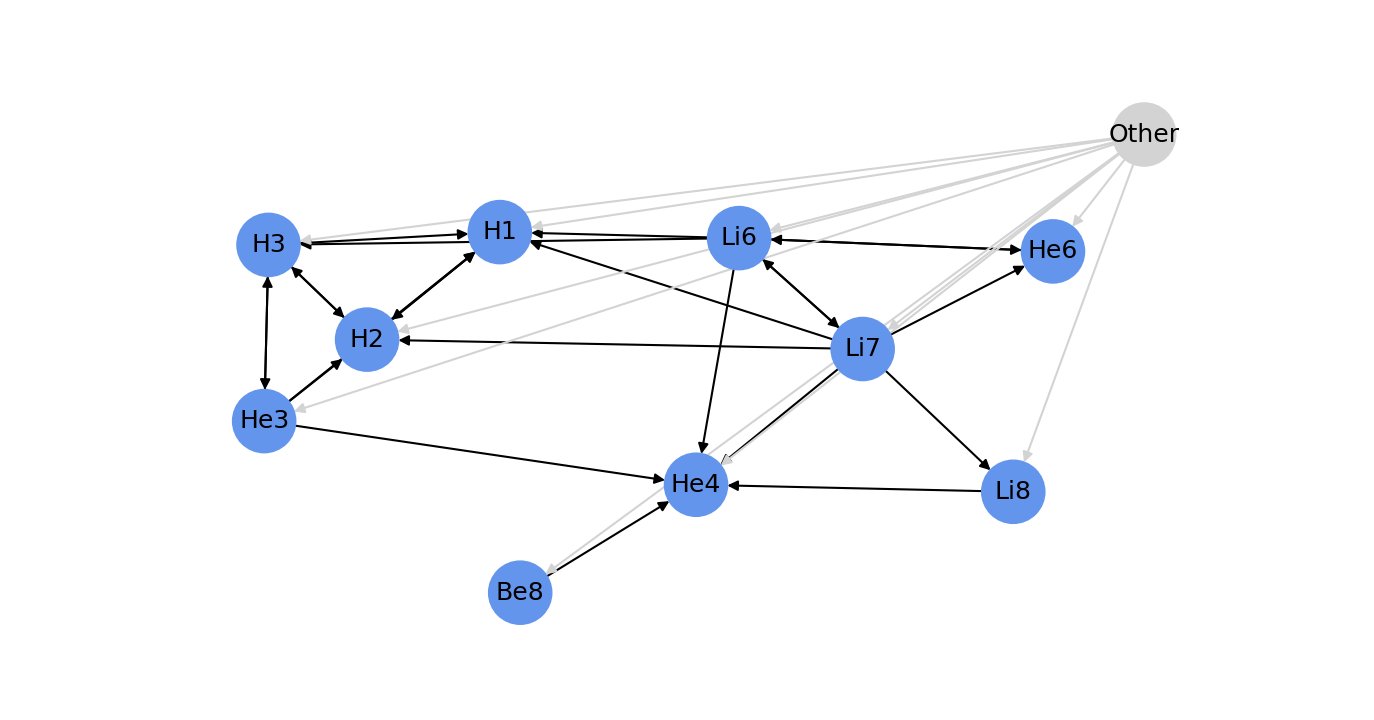}
\caption{The structure of the Hydrogen-Helium-Lithium strongly connected component of the graph. \label{HHeLiComp}}
\end{figure}

From figure \ref{strongComp}, one can see that the subset comprising of Hydrogen, Helium and Lithium appears to form a particular substructure and play an important role in the graph. They indeed seem to be connected to a lot of other isotopes (hence their central position in the representation) in the system and form a strongly connected component. Those light elements are produced in a number of different ways but seem to never contribute to the formation of heavier elements. This can be better seen on Figure \ref{HHeLiComp}, where only light elements are shown, all the other being condensed into a unique "Other" vertex. It is observed that the system formed of Hydrogen, Helium and Lithium form an intertwined network of connections but never contribute to the production of elements heavier than $\,^8$Li. 

\begin{figure}[!ht]
\centering
\includegraphics[scale=0.75]{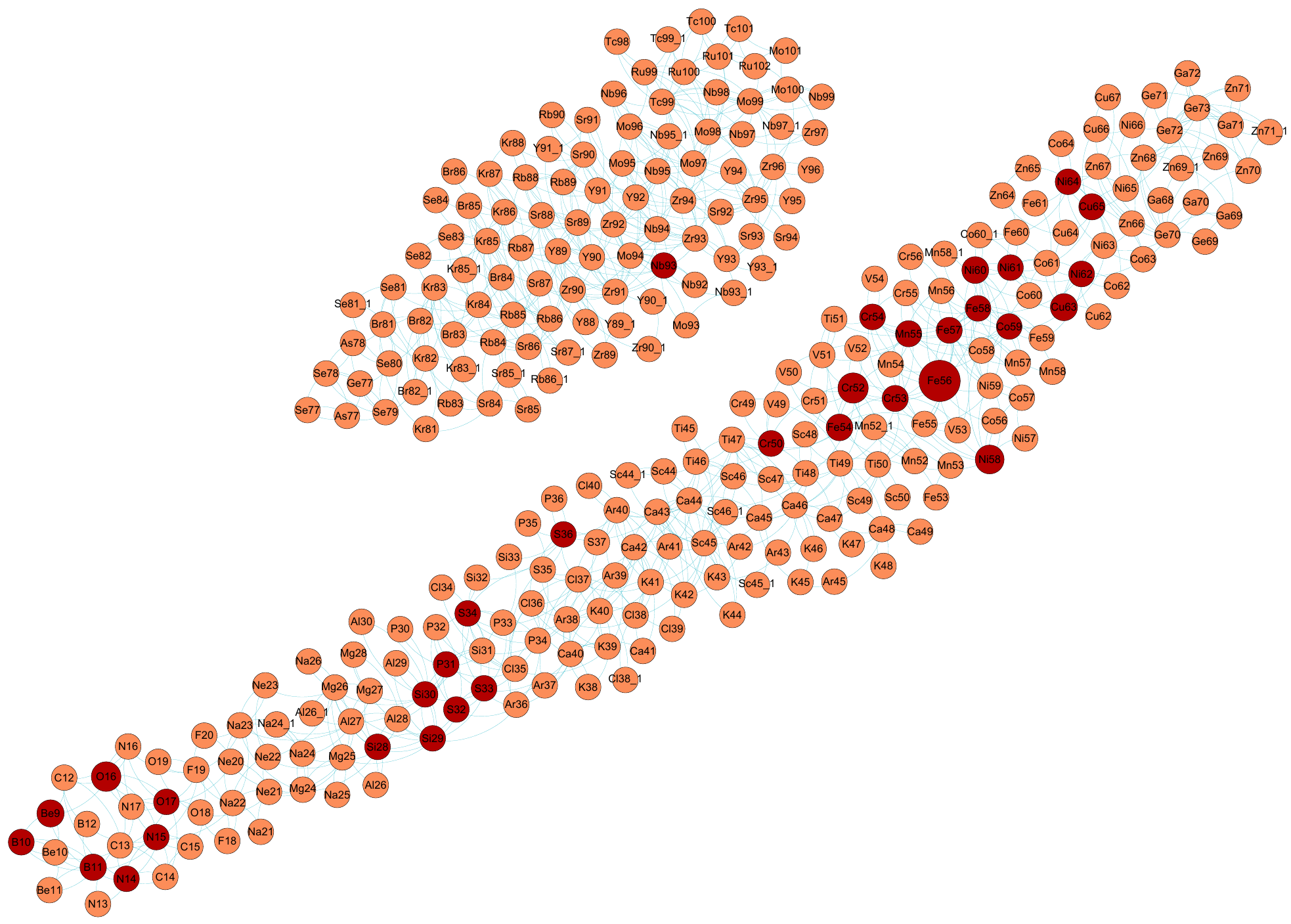}
\caption{The initial graph after removing light elements up to $\,^8$Li. \label{TwoComps}}
\end{figure}

The production of light elements, Tritium in particular, is crucial for a device like ITER and a further study of the structure of the subgraph formed by the light elements might reveal number of interesting physical information. For the sake of the current exercise however, a focus on the production of heavier elements will be favored. In this context, the strongly connected component of light elements does not play any more role in the system and the corresponding vertices and edges might be deleted from the graph. Doing so reveal a drastic simplification in the graph structure, that could have hardly been anticipated: the resulting graph of Figure \ref{TwoComps} is now made of two weakly connected components. 

The explanation for this separation into two components is, here also, perfectly clear from the figure. The upper component of the graph is composed of the ensemble of the daughter nuclides originating from a single initial isotope, namely $\,^{93}$Nb, while the second is mainly driven by the lighter elements of the iron alloy composing the blanket. In five days of irradiation, there has not been time for the nuclear reactions to produce a sufficient number of elements for the two subgraphs to merge. Obviously the two components will not stay separated for the whole irradiation time but if one is only interested in short irradiation period, the separation of the system into two unrelated components is a drastic simplification of the system. 

\subsection{Identiying bridges in the graph}
\label{secBridge}

Related to the notion of connected components of a graph is the notion of \textit{bridges} or cut-edge. A bridge is an edge whose removal increase the number of weakly connected components of the graph. The graph of the ITER model after 5 days of irradiation does contain bridges but only of the most trivial sort: they consist of edges whose removal simply disconnect a single nuclide from the rest of the graph. This simply reflects the fact that there is only one reaction in the system producing this particular nuclide. 

\begin{figure}[!ht]
\centering
\includegraphics[scale=0.75]{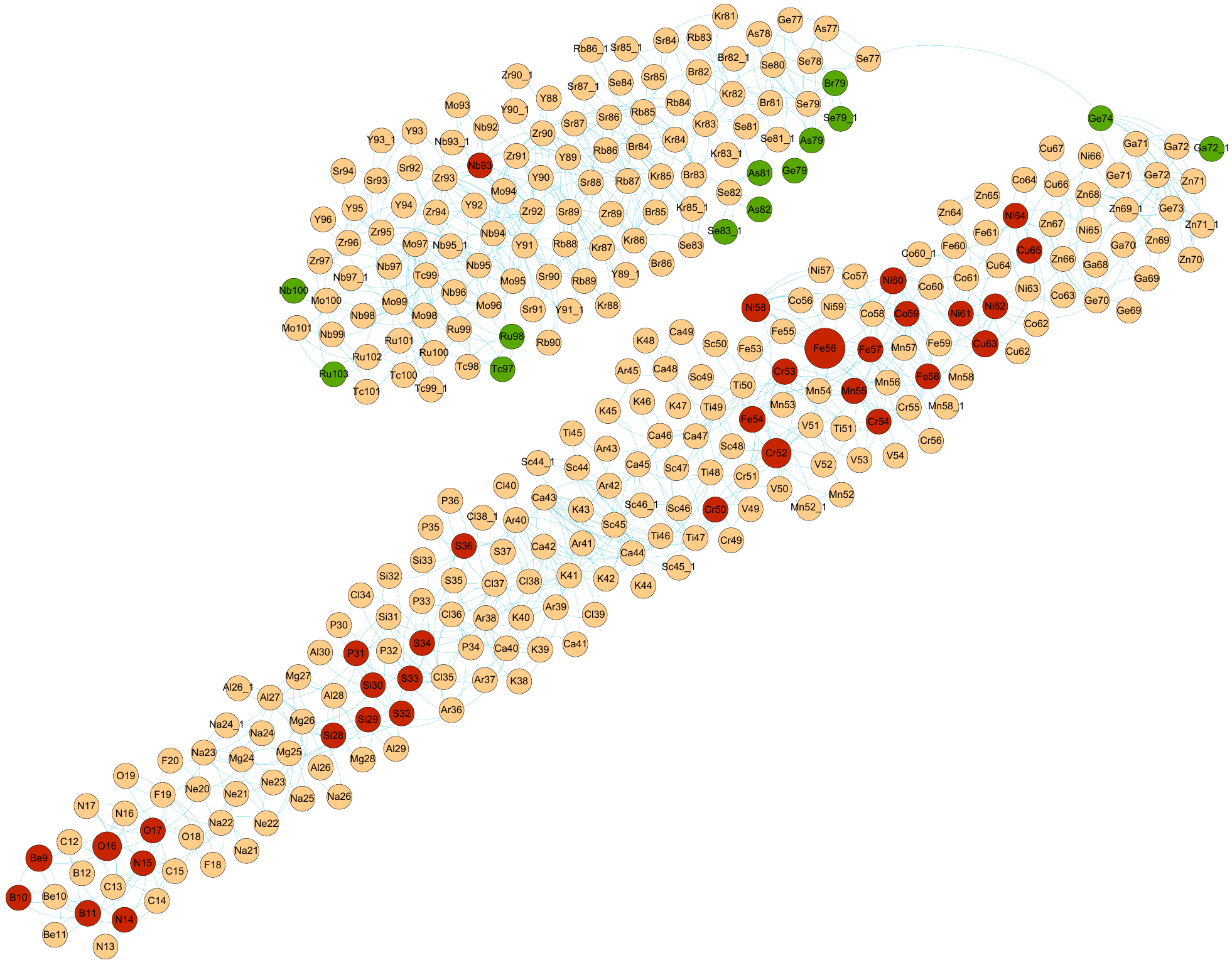}
\caption{Representation of the graph of the system after 15 days of irradiation. New nuclides that appeared in between the two time steps are shown in green.  \label{BridgesTid15}}
\end{figure}

As said before, the graph of Figure \ref{TwoComps} will not stay disconnected for the whole irradiation period. New nuclides will appear in the system and with them, new reactions will slowly introduce more and more connections between the two components. The concept of bridges\footnote{and its generalisation to cut-sets of edges, allowing to search for sets of edges whose removal disconnect the graph.} is then a convenient mean to characterize the appearance of those connections. The state of the system after 15 days of irradiation is presented on Figure \ref{BridgesTid15}. In this situation, the $(n,\alpha)$ nuclear reaction $\,^{77}_{34}Se + n \rightarrow \,^{74}_{32}Ge + \alpha$ is a bridge between the two previously separated components of the graph. Its dynamical behaviour is therefore crucial to understand the interaction between the two pieces that otherwise can be analysed separately.

\subsection{Modularity classes}

\begin{figure}[!ht]
\centering
\includegraphics[scale=0.75]{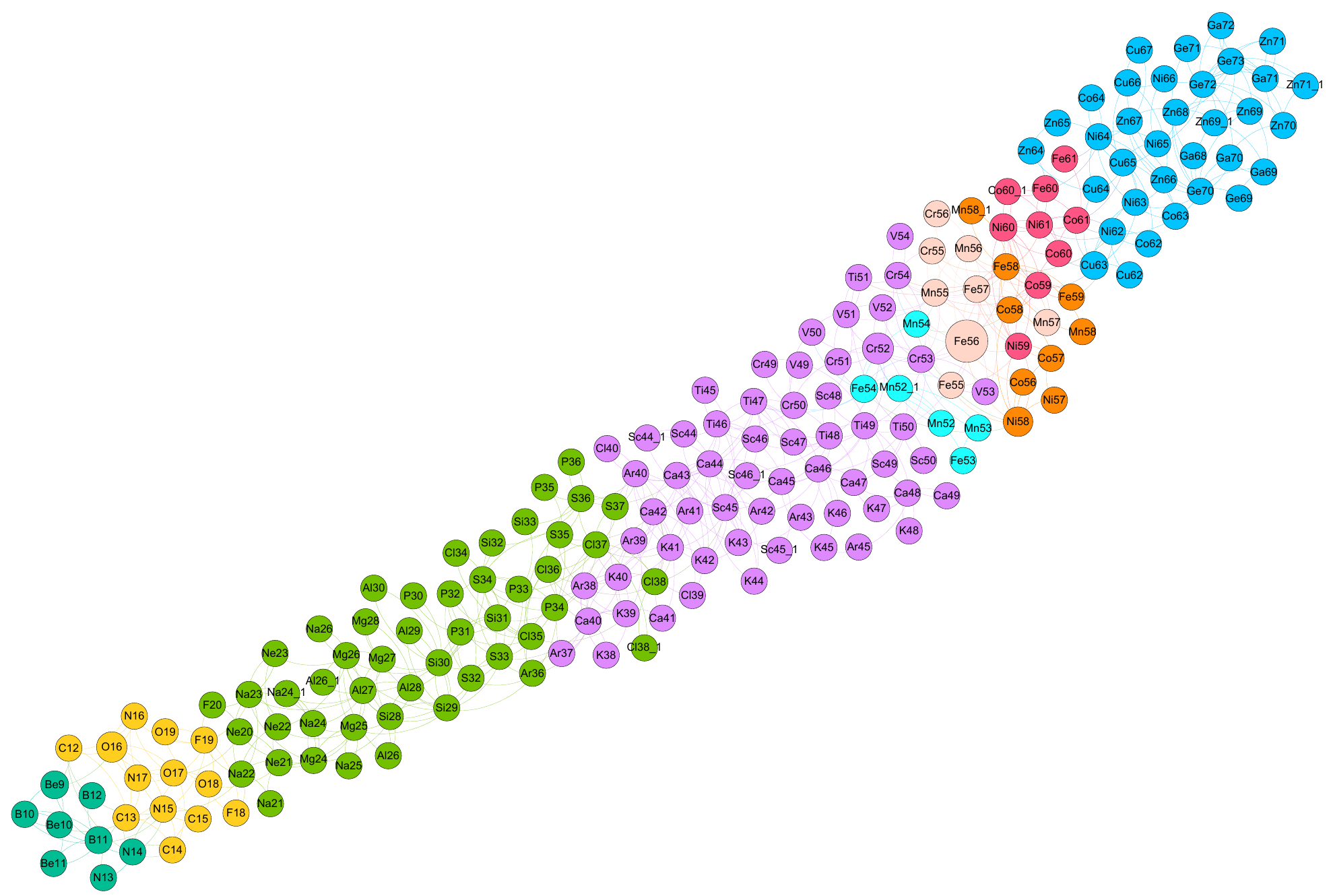}
\caption{Modularity classes built for the iron component of the graph. \label{Modularity_tid5}}
\end{figure}

Returning now to the graph of the system after 5 days of irradiation, it is supposed one is only interested in nuclides belonging to the lower, "iron driven", component of the graph. This component contains 196 nuclides and a bit more than 600 associated nuclear reactions, still much more than what can be easily analysed. 
Nevertheless, there is another topological property of the graph that can be exploited to further cut the system in smaller parts. Indeed, the elongated nature of the component clearly express a physical aspect of nuclear reactions : most reactions occur in between close neighbour, i.e. nuclides sharing similar values of atomic number Z. There is actually many ways in graph theory to formally express this particular idea. In this section, it is implemented using the notion of \textit{modularity classes} \cite{Modularity}. It consists in a heuristic grouping of nuclides being densely interconnected. Figure \ref{Modularity_tid5} shows the modularity class obtained for the iron alloy component of the graph. 

By using modularity classes, one can manage to perform a further subdivision of the problem into smaller bits, which makes the problem much more tractable. The idea is to perform first a separate analysis in each separate modularity class of the system and then take into account the interactions between adjacent classes. By using modularity grouping of the system, one is then assured that the number of interactions in between different groups is reduced to a minimal amount of reactions.



To conclude this section, let us note two important remarks concerning the use of modularity classes for the analysis of a depletion problem. Firstly, it should be kept in mind that the division of the problem in modularity classes is heuristic and does not bear -- a priori -- any particular physical meaning over than a rough estimate of the interactions between regions in the graph. In particular, the division is not unique and applying the same algorithm twice might produce different splitting into classes. 

The second remark concern the applicability of modularity algorithms in problems involving fission. In our particular example, the use of modularity classes appears to be an efficient way to subdivide the system because of the particular elongated shape the graph presents, which is a characteristic of activation problems. In problems involving fission, this particular topology will certainly be lost, because fission add edges between nuclides with very different atomic numbers. In this case, modularity will certainly not be the best tool to analyse the problem and one needs to resort to more advanced or customed methods tailored for this particular problem. In the next section, an example of subdivision that could also be employed when fission is occurring in the system will be provided. 

\section{Simple paths and the reconstruction of decay chains}

Taking advantage of the topological properties of the graph underlying the system, allowed us to subdivide the problem into more manageable pieces. The question now remains to analyze the dynamic structure of the nuclear reactions that occur in each of these parts. To do so, it is proposed to resort to fast and robust methods that exist to navigate along \textit{paths} in the system, i.e. ordered sequence of vertices connected by directed edges. This is achieved by the notion of \text{simple path}, which is a path in a graph that does not contain twice the same vertex (the path does not pass twice at the same place). 
The ability to efficiently navigate along paths in a graph, using well known, available methods \cite{SimplePaths}, is one of the most convenient tool that graph theory has to offer. It will allow to reconstruct in an efficient and robust manner all of the reaction chains occurring in the system. 

\subsection{Identifying the initial nuclides for the production of a given nuclide}

In the present case, where one is interested in the behaviour of the evolved composition at early times, one question that naturally arises is to identify, for each nuclide produced during the irradiation, the initial isotopes that contributed the most to its production. In other word, one would like to associate to each nuclide a notion of how likely it was produced following a chain of reactions that originates from a reaction of one of the initial isotopes composing the material. In practice, the association is constructed by following all of the simple paths from the initial isotopes $N^{T=0}_i$ to a nuclide of interest $N$.

Using the well known algorithms of graph theory, it is straightforward to obtain all of the simple paths from one vertex to another (path finding algorithms achieve $o(V+E)$ time). In the application sought, the key point is to assign a probability to each individual path from $N^{T=0}_i$ to $N$. To do this, one needs to have a good view of the system and adapt the notion of probability so as to match the question asked about the system.

\begin{figure}[!ht]
\centering
\includegraphics[scale=0.5]{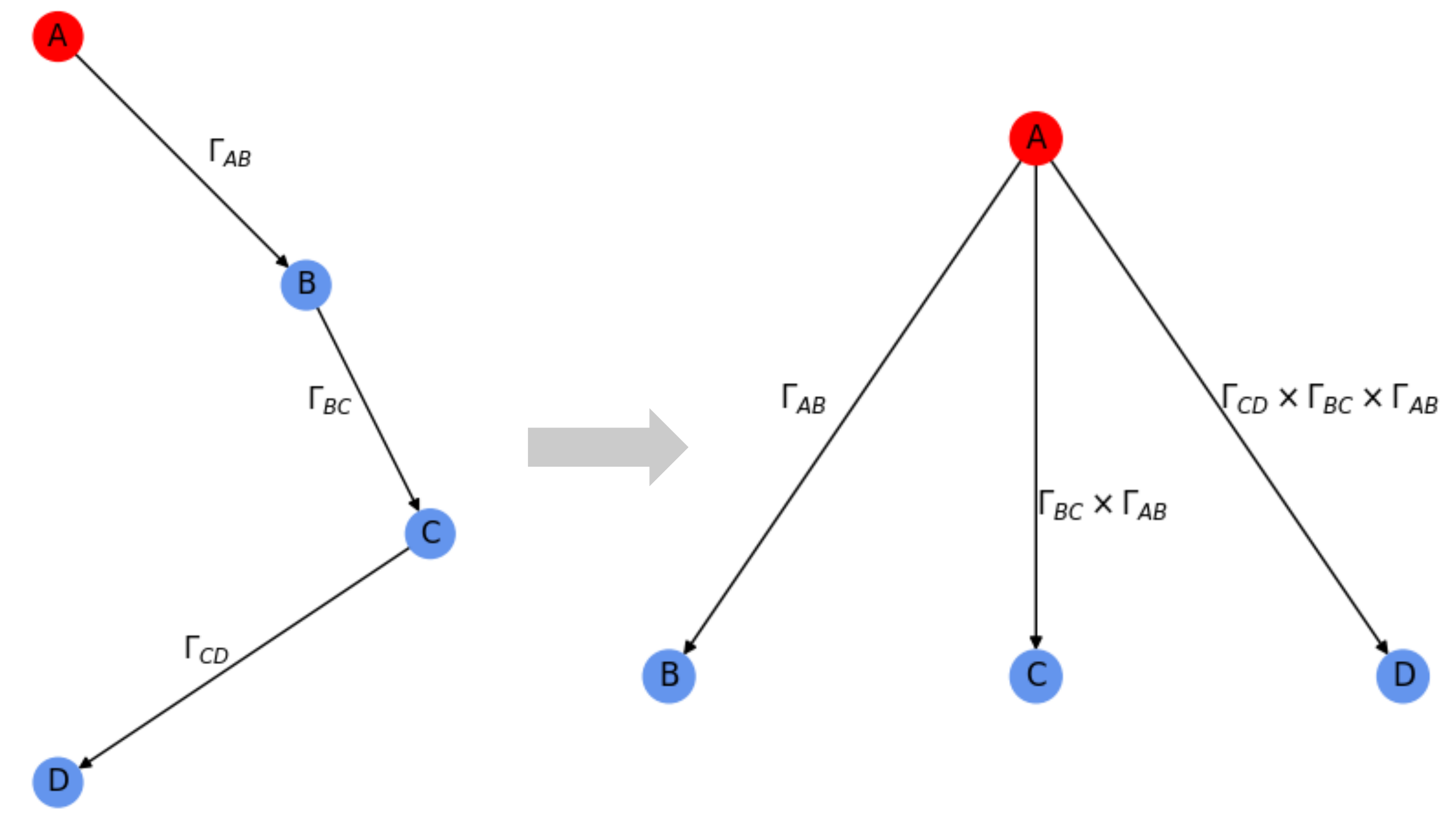}
\caption{Following the paths from a source isotope $A$ to a target isotope $D$, we can associate a probability that the production of $D$ is due to the isotope $A$, by carefully multiplying and summing branching ratios. \label{LinkToInitalIsotopes}}
\end{figure}

The most simple solution one could think of would be to associate each edge in the path with the branching ratio of the reaction it encodes. For one path from $N^{T=0}_i$ to $N$, we could then multiply the branching ratio of the edges composing the path, as illustrated on Figure \ref{LinkToInitalIsotopes}. The numbers obtained could then be interpreted as the probability to end up in vertex $N$ by following the reaction chain along a path starting from the isotope $N^{T=0}_i$. Summing over all paths from  $N^{T=0}_i$ to $N$, one can then obtain the total probability to produce some nuclide $N$ starting from a position $N^{T=0}_i$ in the graph. 

The method just built is valid but does not really answer the question raised in the beginning of this section, which was to identify, for a given nuclide $N$, the initial isotope it most likely originate from. To answer this question, one has to follow the paths backward and thus associate to each path a notion of "ingoing" branching ratio. This notion can be defined in a simple manner, by the ratio of the rate of the reaction one is interested in (i.e. the reaction on the path under consideration) to the sum of the reactions rates ending on the nuclide. Multiplying the ingoing branching ratio on each path from $N^{T=0}_i$ to $N$, one obtains a measure whose interpretation can be thought of as the likelihood of a nuclide $N$ to originate from the initial isotope $N^{T=0}_i$.

\begin{figure}[!ht]
\centering
\includegraphics[scale=0.75]{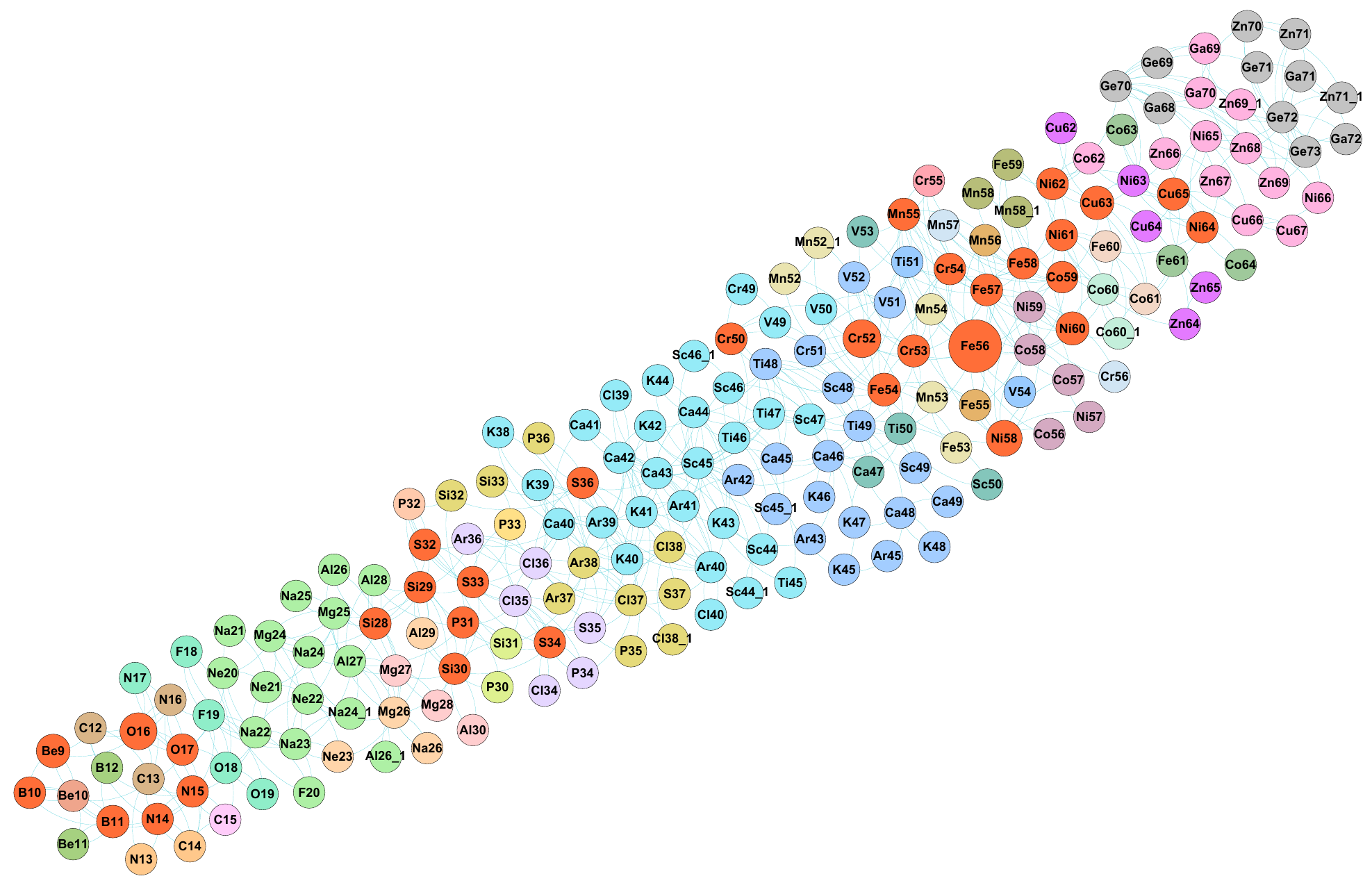}
\caption{Each nuclide on this figure is attributed a unique color depending on the specific initial isotope it most likely originate from. (Initial isotopes are colored in red). \label{LinkToInitalIsotopes_allIsotopes}}
\end{figure}

Performing the computation for all pairs $(N^{T=0}_i, N)$, we can use the probabilities to associate each nuclide in the system with a unique (or at least reduced number of) initial nuclide whose nuclear reactions is most likely to be responsible for its creation. In our fusion activation example, such association was performed for the iron component of the evolved composition after five days of irradiation and is illustrated on Figure \ref{LinkToInitalIsotopes_allIsotopes}. Each nuclide in the system is given a unique color which is associated to the initial isotope that most likely was responsible for its creation. 

Figure \ref{LinkToInitalIsotopes_allIsotopes} gives an approximate but enlightening idea of what happened in the system during the first few days of irradiation. One can see for instance, that all of the nuclides in blue in the middle of the graph originate from the same initial nuclides (Chromium 50 and 52) and form an intertwined web of rapid successive nuclear reactions connecting ultimately with the less massive elements of the graph. If one is particularly interested in the production of one of the elements in this region, it can be directly understood how it is influenced by the initial composition of Chromium. Thus, one gets a rough idea of the decay channels that are involved. It can then be used to further simplify the system and allow for more specific treatments. 

Another comment can be made about the resemblance of Figure \ref{LinkToInitalIsotopes_allIsotopes} with the modularity classes found on Figure \ref{Modularity_tid5}. It appears quite natural that modularity classes, in identifying groups of tightly bonded elements, is able to broadly perform an association between nuclides and an ensemble of initially present isotopes. Thus the coloring of Figure \ref{Modularity_tid5} must prefigure the results obtained by a much finer association, such as the one presented on Figure \ref{LinkToInitalIsotopes_allIsotopes}.

\begin{figure}[!ht]
\centering
\includegraphics[scale=0.6]{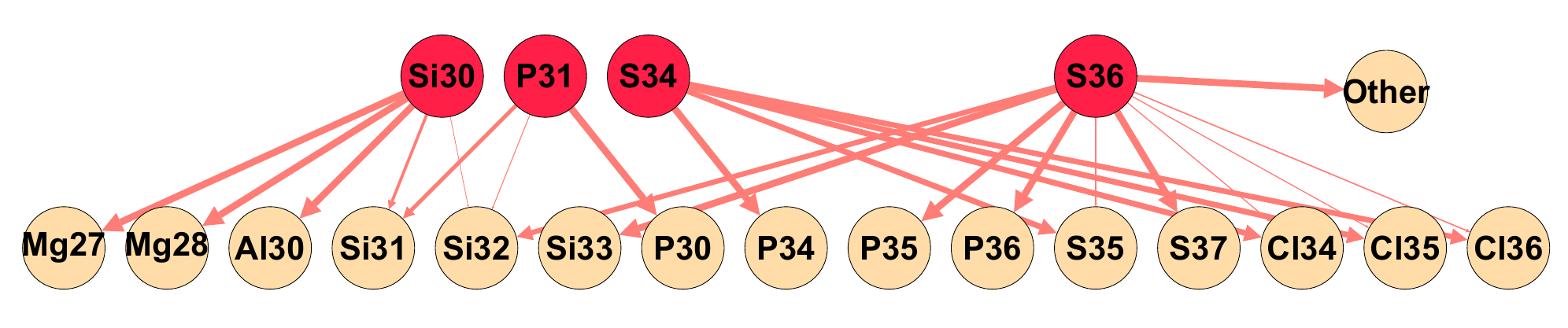}
\caption{An extract of the bipartite graph constructed from the association of one isotope initially present to the isotopes that appeared during the irradiation. \label{Bipartite}}
\end{figure}

To end this section, note that by associating the nuclides that appeared in the system with isotopes that were initially present, one is effectively building a particular kind of graph -- a \textit{bipartite graph} -- were edges always start in one family of vertices (the initial isotopes) and point to all of the other nuclides. This type of graph can also be an invaluable tool to visually display and understand the dynamics of the activation problem. An extract of the bipartite graph, constructed from the association explained in this section, is presented on Figure \ref{Bipartite}.

Let us state finally that the pairing that was made in this section must be taken with a bit of caution and do not aim at being something else than an heuristic mean to analyze in a simplified fashion the nuclear reaction chains. Some more elaborate methods could be put in place, in particular to take into account the timescales associated to each nuclear reactions taking place and thus to gain more insight on the dynamics of the system. 

It is nevertheless believed that the method devised in this section gives a good idea of what is happening in the system. It can be used to further subdivide the graph, allowing for the much finer analysis that is proposed in the next sections. 

\subsection{Construction and visualisation of the nuclear reaction chains}

Following the subdivision of the problem into smaller, more manageable pieces, the next step of the analysis would consist of analysing specific behaviour of substructures of the graph, emphasizing on the small groups of nuclides who are playing a predominant roles for the specific applications sought. We will not delve here in a detailed study of the precise understanding of the origin of the structures that we previously found in the graph, because that would take us too far from the main subject of this paper. We rather take advantage of this section to present the last types of graph that we can build from nuclear reaction chains, \textit{tree graphs}.  

\begin{figure}[!ht]
\begin{subfigure}{.5\textwidth}
\centering
\includegraphics[scale=0.55]{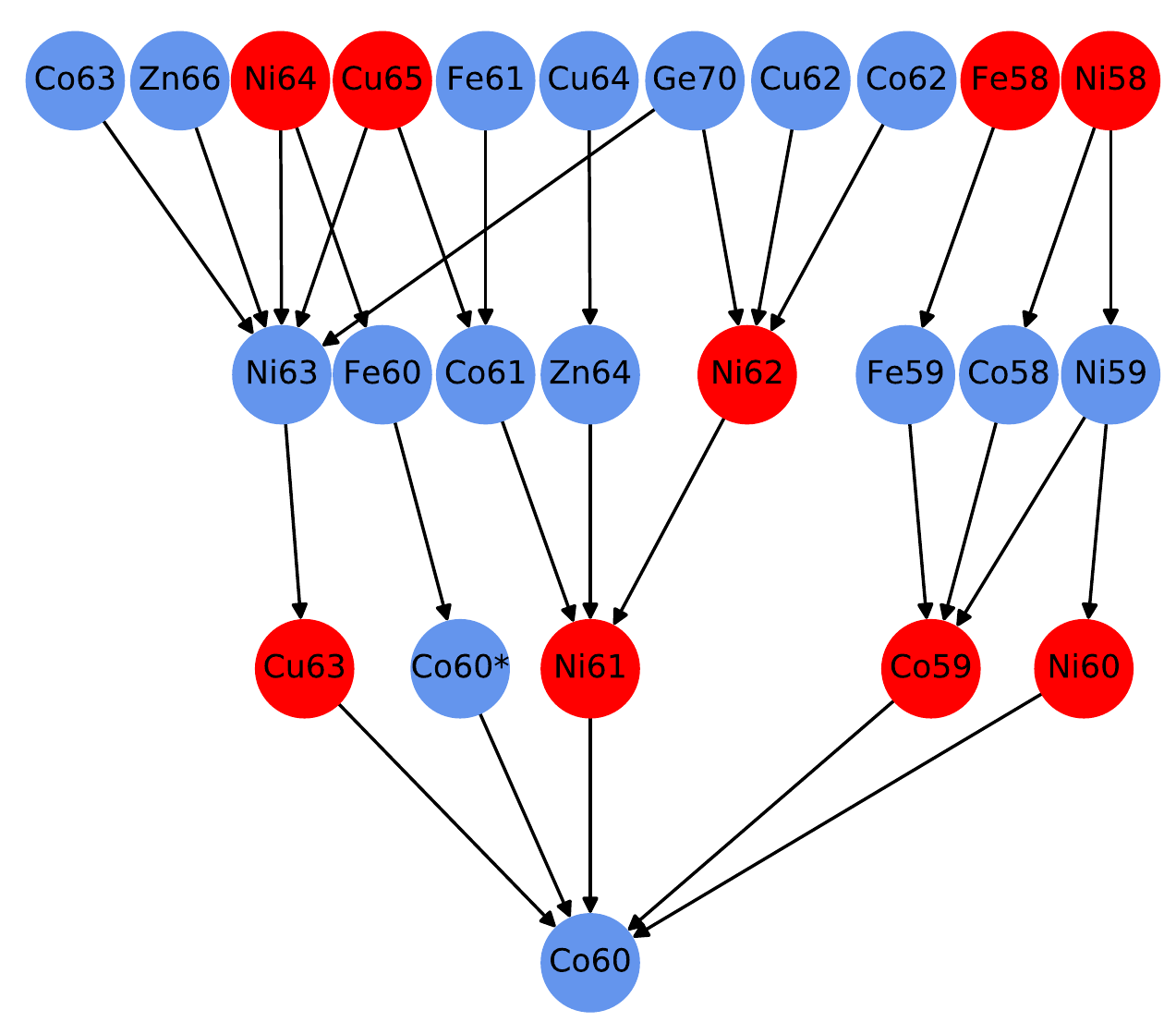}
\caption{Tree graph of ${}^{60}$Co production. \label{ProdCo60}}
\end{subfigure}
\hspace{0.1 cm}
\begin{subfigure}{.5\textwidth}
\centering
\includegraphics[scale=0.55]{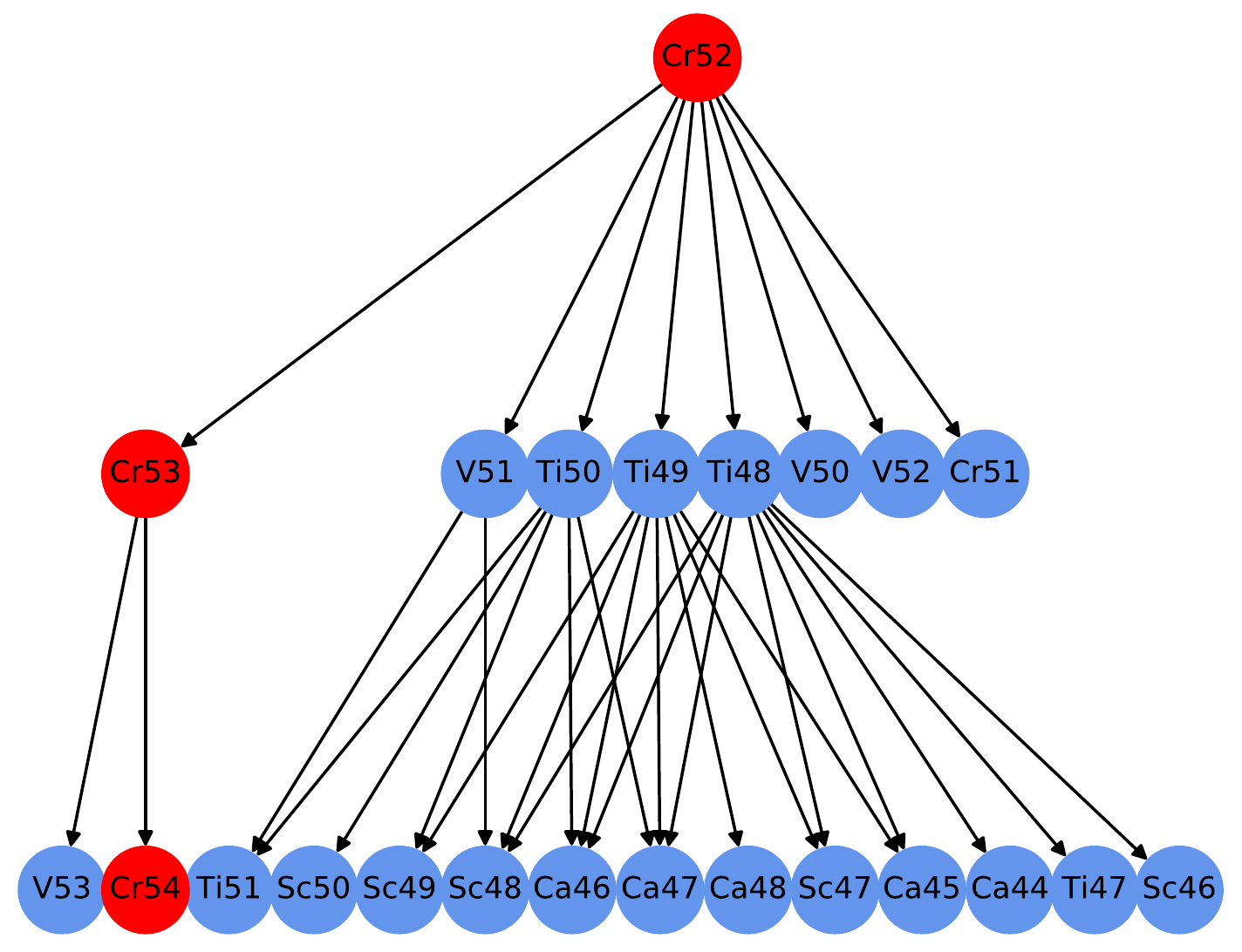}
\caption{Tree graph of ${}^{52}$Cr disintegration.\label{DecaysCr50}}
\end{subfigure}
\caption{Two examples of tree graphs. (a) The production channels of ${}^{60}$Co. (b) The disintegration channels of ${}^{52}$Cr.\label{DecayTree}}
\end{figure}

Representing nuclear reaction chains as a tree-like structure is perhaps the most practical way to analyze the structure of the production and decay channels of a specific nuclide. Two examples are shown on Figure \ref{DecayTree}, presenting the production channels associated to the appearance of ${}^{60}$Co and the disintegration chain of ${}^{52}$Cr, for which it was observed in the last section that it occupy an important place in the iron component of the total graph. 

Although this kind of graph has much to offer for a precise survey of the structure of the nuclear reaction chains, it is in general not a structure that naturally arise in the entangled web of nuclear reactions. This is because, nuclear reaction chains tends to form cycles, that is to say paths which follow closed loops, passing several times through the same position. For this reason, a specific methodology must be engineered to bring out this type of structure from the total graph of the system. The most practical method is based on the exploitation of simple paths between nuclides, which, because the paths are simple, naturally do not include any cycle. This notion of simple path directly allow us to extract the tree-like that was sought. 

\section{Studying differences in between two graphs}

Because graphs are well defined mathematical objects, simple algebraic operations like addition and subtraction in between two weighted graphs can be given a definite and precise meaning. This is yet another major advantage of graph theory which comes in handy in situations where one wants to compare two systems sharing common properties or irradiation histories. This can happen for instance if when one is simulating the same system, using two different sets of nuclear data. It can also be an invaluable tool to analyse how the system is evolving in time. 

In this last section, it is proposed to highlight some of the features one can extract from a graph using simple vertex-wise and edge-wise operations. Those methods will be applied to the ITER activation model, in both of the above mentioned cases, namely the comparison of nuclear data libraries and the comparison of the system at different time steps.  

\subsection{Identifying differences between nuclear data libraries}

The results presented thus far were obtained using the JEFF-3.1 nuclear data library. In this section, a comparison of those results with a simulation of the ITER model based on the ENDF/B-VII.1 nuclear data library \cite{endfb71} will be made. For the activated material after five days of neutron irradiation, one ends up with two graphs that can be compared.  

\begin{figure}[!ht]
\centering
\includegraphics[scale=0.7]{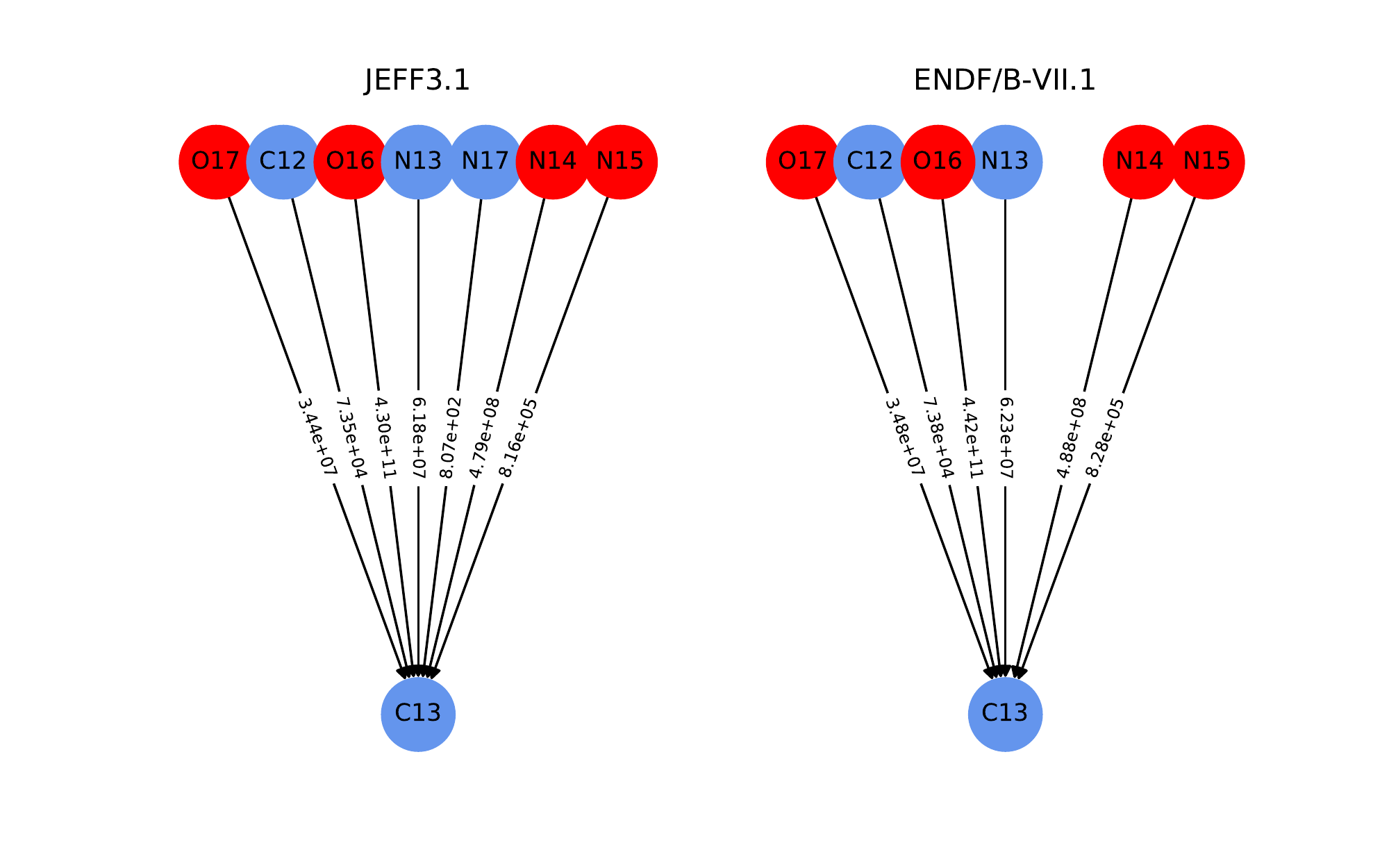}
\caption{The production tree of ${}^{13}$C, shown for the graph constructed from JEFF3.1 (left) and ENDF/B-VII.1 (right). Total reaction rates (in $s^{-1}$) are displayed for each reaction.  \label{DecayDiff}}
\end{figure}

The first thing one could do, is to check if the two graph contains the same number of nodes and edges, that is to say, verify if the irradiation made using two distinct nuclear data libraries involves the same number of nuclides and nuclear reactions. In the present case, the two graphs involve the same number of nuclides but there is one particular reaction that is missing from ENDF/B-VII.1, it is the $\beta^- + \alpha$ reaction of ${}^{17}$N, decaying into ${}^{13}$C, see Figure \ref{DecayDiff}.

On can see from Figure \ref{DecayDiff} that the reaction rate associated with this missing reaction is much lower than any of the other production channel of ${}^{13}$C. One can also show that this channel is also far from being a dominant reaction in the decay of ${}^{17}$N. So it is concluded that no major discrepancy will be observed because of the absence of this particular reaction in ENDF/B-VII.1. 

\begin{figure}[!ht]
\centering
\includegraphics[scale=0.8]{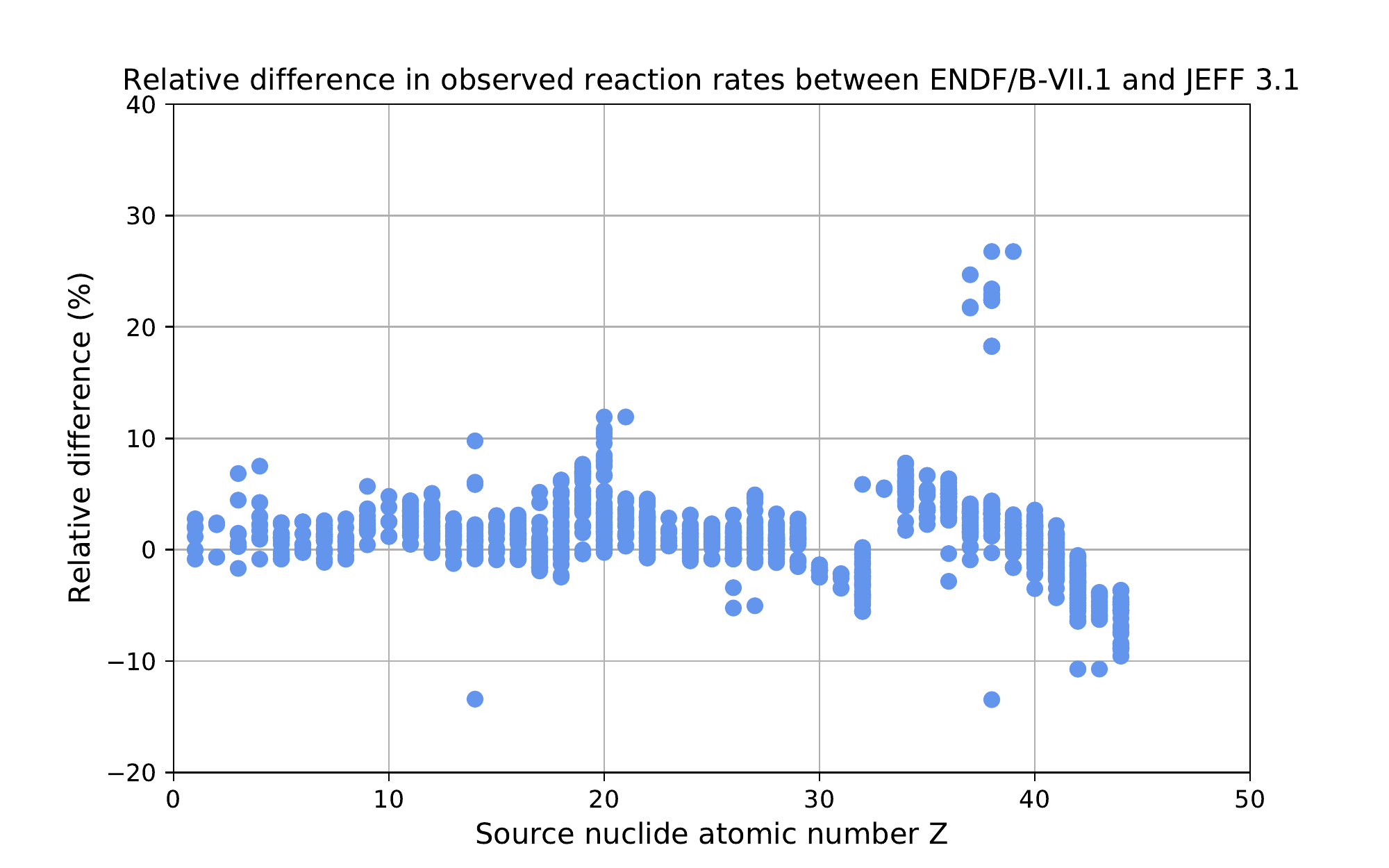}
\caption{Relative differences (ENDF/B-VII.1 - JEFF 3.1)/ENDFB-VII.1 obtained by comparing the instantaneous reactions rates of the fusion activation problem after five days of irradiation. The abscissa represent the atomic number of the nuclide responsible of the reaction. \label{RelDiffTot}}
\end{figure}

For all of the other reactions that the two nuclear data libraries have in common, we can plot their relative difference to survey the overall relative behaviour of both data sets. These relative differences are presented on Figure \ref{RelDiffTot}. 

Note that each point of the graph represent the total disintegration rate for each of nuclides in the system. This rate is proportional to the isotopic concentration of each nuclide "source" of the associated reaction. This remark being said, it should be understood that the highest discrepancies obtained in the graph does not reflect problems in nuclear data but are mainly obtained for short lived nuclides that are produced in very small quantities in the system. Thus, an overall a good agreement between the two nuclear data libraries is observed. 

\subsection{Tracking the temporal evolution of the system}

Finally, the temporal evolution of the system during the irradiation period is discussed. From a practical point of view, the temporal evolution is given as a set of graphs, one for each time step at the output of the depletion code. Simple manipulations on the set of graphs can be applied to extract useful information about the temporal behaviour of the system.

\begin{figure}[!ht]
\centering
\includegraphics[scale=0.6]{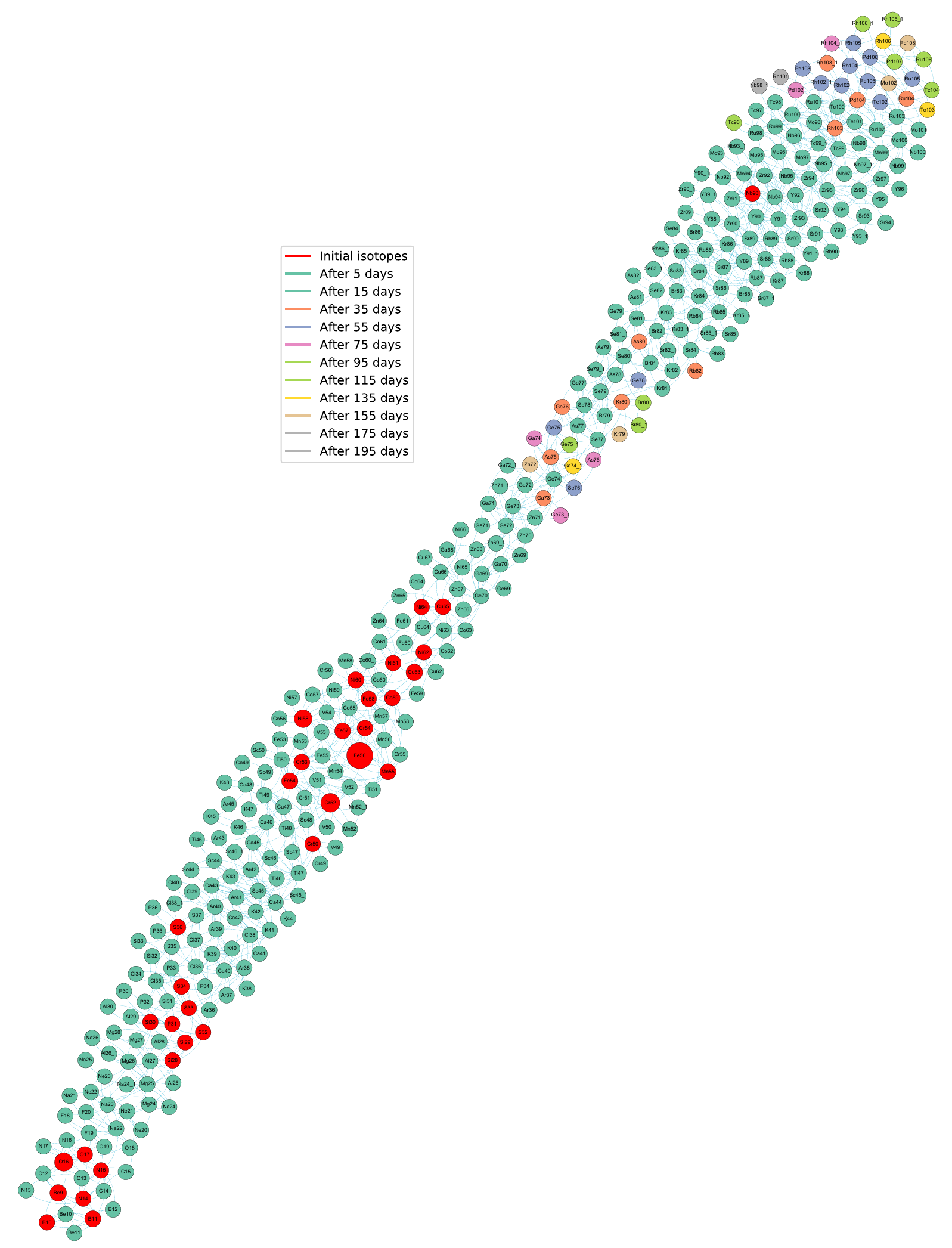}
\caption{The graph of the system (without the Hydrogen-Helium-Lithium component) after 195 days of irradiation. Each isotope is colored based on the time step at which it first appeared. \label{Evo}}
\end{figure}

In doing so, the effort spent understanding the structure of the system during the first moments of the irradiation history will not go wasted. For now one can use the knowledge accumulated and examine each new structure in light of the basic shapes that are by now familiar. As an illustrative example, Figure \ref{Evo} show the graph of the system after 195 days of irradiation (like before, the Hydrogen-Helium-Lithium structure was suppressed, to gain visibility). As it was mentioned back in section \ref{secBridge}, the graph is now made into one, weakly connected component. The isotopes that appeared along the evolution are represented using different colors. On overall, it can be seen that, at least for this simple activation example, the material's composition is still built around the basic outline of the first few days, and is little by little enriched with new isotopes. New isotopes rapidly fill in the gap between the initially disconnected components. Once this is done, the only additional effect of irradiation is the creation of heavy elements, that continue to appear even at the very end of the irradiation period. 

To conclude this paper, let us stress that by looking only at the graph of Figure \ref{Evo}, but this is also true for any other figure of the paper, one is in reality putting aside some of the most important dynamical aspects of the system. Indeed, the concentration and reactions rates are constantly evolving in the system and the figure does not seem to be able to capture this aspect of the temporal evolution. To understand the true dynamical behaviour of the system, one must resort to a detailed analysis of the composition of each isotopes, in the same fashion that one would do if one had not had graph theory tools at disposal. 

It seems like graph theory is not capable of capturing these aspects of the evolving system. But this is merely due to our emphasizing of visual aspects linked with graph representation of the system. In providing a robust and convenient structure to navigate through, graph representation also is a most efficient container to use in a detailed analysis of more quantitative aspects of a depletion problem. Using modern and efficient frameworks, such as the NetworkX Python package, one is left in a position to grasps many more information of a depletion problem.



\section{Conclusion}

A broad overview of the most commonly known methods used in graph theory has been achieved and their applicability to the analysis of a depletion problem has been demonstrated. Graph theory indeed is a natural representation to visualize an irradiation process but it is also the right tool to disentangle its associated complex web of nuclear reactions and extract from it sensible and useful information, that would otherwise remain out of reach. 

Our attention was focused on some of the most simple and readily available methods that a user could use to understand the global structure underlying the activation problem that was proposed. The few solutions that have been presented, will figure among the methods that will, in term, be made directly available to users of the AURORA analysis software, for the post-processing of the outputs of the VESTA Monte-Carlo depletion code. 
To keep the directing line of the paper at the level of a broad overview and to emphasize on the methods that AURORA will natively be provided with, the proposed exploration of graph theory presented in this paper is only partial. Much more tools and methods comprised within graph theory could still be applied and help to further understand the delicate problem of nuclear disintegration chains. 

Many more tools inspired by graph theory could be used to gain meaningful insights about depletion problems. For instance, one could use metrics or indices (\textit{centrality measures} for example) to identify in a given system the isotopes or nuclear reactions that are the most important, either because they drive the production of important groups of nuclides or because they induce reactions whose uncertainty could cause a major problem for estimating the composition of other isotopes. Over methods and algorithms could also include the notions of quotient graphs, multi-commodity flows, the study of discrete optimal transport on the graph and the emerging field of graph neural networks, that could reveal itself to be of great interest either to extract sensible information from a graph or construct fast and efficient ersatz of a depletion problem, to predict nuclear inventories in situations where rapidity takes precedence over accuracy.   

\pagebreak
\section*{Acknowledgments} 

The author would like to thank Mariya Brovchenko for helpful discussions and comments.

\pagebreak
\bibliographystyle{ans_js} 
\bibliography{bibliography}

\end{document}